# Needs-guided engineering for adolescent mental health

**Qingqi Zhang[1] · Yaqun Wang[2] · Tianbin Hu[3] · Ning Ma[1,4]*

[1] Hangzhou Institute of Medicine, Chinese Academy of Sciences, Hangzhou 310018, Zhejiang, China

[2] Tongde Hospital of Zhejiang Province, Hangzhou 310012, Zhejiang, China

[3] Zhejiang Provincial People's Hospital (Affiliated People's Hospital), Hangzhou Medical College, Hangzhou 310014, Zhejiang, China

[4] Research Center for Psychological Sciences, Zhejiang University, Hangzhou 310058, Zhejiang, China

Correspondence to Ning Ma: maning@him.cas.cn

## Abstract

Adolescent mental health support requires decisions that begin with developmental needs, not devices. This Review integrates psychological mechanisms, evidence from established interventions, comparative technology selection, product architecture and real-world validation into a needs-guided framework. Using Chinese adolescent contexts, we illustrate how similar participation difficulties warrant different responses, from school-level action to non-immersive care or conditional virtual reality, and define evidence gates for demonstrating incremental value.

## Introduction

An adolescent who rarely speaks in class may fear negative evaluation, have few opportunities to practise conversation, lack sleep, or face intimidation by classmates. Each pathway can produce a similar visible outcome while calling for a different response. A system that measures speaking time and rewards longer answers may therefore improve an easily observed performance without addressing the problem that restricts the young person's participation. The question that should precede the design of any device is not what it can detect or simulate, but which developmental opportunity is constrained, why it is constrained and what change would matter in everyday life.

Mental health conditions contribute substantially to the burden of ill health in adolescence, but adolescents also experience functional difficulties below diagnostic thresholds, and ordinary developmental variation must not be pathologized [1]. Autonomy, competence and relatedness are established basic psychological needs in self-determination theory [2]. Broader developmental accounts also emphasize safety, identity exploration and participation within the opportunities offered by family, school and wider society [3]. These are different kinds of constructs: basic psychological needs are not interchangeable with symptom scales, school grades, use time or a clinician's diagnostic category. Needs provide the justification for intervention; symptoms and behavioural observations help locate the barriers to meeting them.

At the same time, adolescent support is an increasingly engineered activity. Remote cognitive behavioural therapy (CBT), classroom programmes, smartphone tools, automated dialogue and immersive virtual environments provide different ways of delivering information, generating experience, recording decisions and supporting behaviour change. Engineering expands the available design space. It does not, by itself, identify an intervention target or establish therapeutic superiority. Recent arguments for purpose-driven digital mental health and systems thinking in bioengineering likewise call for aligning technologies with real-world objectives, contextual constraints and verification requirements [4, 5]. An important next step is to make those principles operational across disciplines.

Here we synthesize developmental and contextual evidence, psychological models, comparative intervention research, and engineering methods into a six-decision framework: needs, mechanisms, evidence, technology, product and validation (FIG. 1). The framework is not a new causal law or a scoring scale. It is a decision architecture: each transition makes assumptions explicit, names the evidence needed and provides a route to revise or reject the proposed solution. China offers an informative case because adolescents experience differentiated school demands, family relationships, peer expectations and access to digital infrastructure. The resulting design principles, however, can be evaluated in other societies and health-service settings. We appraise evidence by distinguishing observational associations, mechanistic experiments, controlled intervention outcomes and independent everyday functioning. Comparator strength, participant age and clinical status, measurement validity, implementation burden and generalisability govern how findings inform technology choice. The synthesis is critical and narrative, organized by functional targets rather than by device categories.

## Developmental needs define the problem

### Needs are not equivalent to performance targets

A need is a reason for providing support; a target is a function or mechanism that an intervention can plausibly change. A young person may need greater social belonging, yet the observable concern reported by school staff may be infrequent participation in group tasks. Increasing participation count is useful only if the change is wanted by the adolescent, appropriate to the classroom and linked to the intended social or developmental benefit. More speech is not necessarily better speech, and greater compliance is not necessarily greater autonomy. Similarly, an improved laboratory score may measure adaptation to test demands rather than a lasting increase in self-directed functioning.

This distinction is especially important during adolescence, when growing independence coexists with unequal access to resources and strong dependence on peers and adults. The same exercise may support competence for one adolescent and threaten agency for another. Design studies should therefore specify a target population, the activity currently limited, a youth-relevant objective, and competing explanations for the observed restriction. Selection of a clinical outcome should be justified by clinical status. For non-clinical support, indicators of voluntary participation, self-directed action, school engagement and perceived capability may be more relevant, but each requires a valid measurement procedure.

### Social ecology changes the expression of needs

Developmental needs do not change every time a new technology appears; the opportunities and constraints under which they are satisfied do. A three-wave study of 2,965 adolescents in Zhejiang, China linked academic stress to later depressive and anxiety symptoms and identified sleep and loneliness as statistical mediators [6]. This provides evidence for linked processes, not proof that any single pathway caused the observed emotional difficulties. A separate longitudinal study of 996 Chinese adolescents identified joint developmental trajectories of parental psychological control and autonomy support associated with adjustment [7]. Such results make it insufficient to treat family relations or academic demands as interchangeable background variables.

Digital environments add further settings in which adolescents experience feedback, comparison, friendship and exclusion. Research comparing adolescents with and without mental health conditions has identified differences in social-media experiences, including social comparison and the effects of feedback on mood [8]. These observations call for more precise characterization of users and contexts; they do not support a single-cause account in which digital media uniformly produces anxiety or depression. Exposure, response style, clinical status, school opportunity and family context may each modify the relationship. The methodological implication is to measure environment and development alongside symptoms, instead of treating them as noise to be eliminated by a personalization algorithm.

Population averages also conceal the context of individual decisions. School policies, rural–urban differences, the availability of clinicians and student support staff, families' ability to afford equipment, and norms concerning classroom expression affect which intervention can be implemented fairly. A device that performs consistently in a laboratory may have a different benefit–burden balance in a crowded school or in a household without private space. This is why needs and contextual assessment must precede technology selection, and why the China-specific evidence is used as a test of the framework's sensitivity to heterogeneous conditions rather than as a cultural stereotype (BOX 1).

## Mechanisms specify what can change

### One difficulty, several pathways

Needs alone do not define an intervention component. Figure 2 illustrates how reduced classroom participation can arise through at least five candidate pathways: anticipated negative evaluation and avoidance; limited opportunities to practise social communication; stress- and sleep-related reductions in available self-regulatory resources; perceived lack of agency or efficacy; and exclusion or an unsafe peer environment (FIG. 2). These pathways can coexist. They cannot be inferred reliably from a single behavioural trace, and their relative contribution may change across developmental stages and settings.

For social-evaluative anxiety, a useful mechanism-focused formulation links threat expectation, anticipatory distress, safety behaviours and avoidance to reduced opportunities for corrective learning. The inhibitory-learning approach to exposure therapy describes how disconfirming expected negative outcomes can be more informative than immediate anxiety reduction alone [9]. This motivates experimental manipulation of social feedback and expected consequences, but it does not establish that an immersive simulation is always the best place to do so. In some circumstances, guided in-person participation may supply more meaningful disconfirmation and better transfer.

For communication difficulties, the relevant limiting factor may instead be knowledge of conversational conventions, pragmatic skill, practice opportunities or supportive peer responses. Teaching a repertoire of actions, providing realistic feedback and changing the interaction context may be more appropriate than repeating exposure. Difficulties associated with sleep, persistent stress or overloaded schedules similarly point toward daily routines and conditions that generate the burden. Under peer victimization, the primary target may lie outside the individual entirely; training a young person to tolerate an unsafe situation without changing it risks displacing responsibility from institutions to children. A meta-analysis of 69 randomized school anti-bullying trials involving 111,659 participants found small but meaningful group-level effects, supporting school-level action as one directly relevant evidence source [10].

### From correlate to validated target

Mechanistic claims require progressively stronger evidence. Correlations between head movement, autonomic signals or questionnaire scores and anxiety identify possible markers, not necessarily mechanisms. Longitudinal association improves temporal characterization but may remain confounded. Experimental manipulation can establish that a target is responsive to an intervention component; establishing that change in the target causes improvement in independent outcomes requires stronger mediation or factorial evidence. A mechanism-focused product specification should state (i) the construct and its operational definition, (ii) the manipulation predicted to affect it, (iii) the process measure, (iv) the temporal ordering expected, and (v) the independent functional outcome.

The distinction also limits the interpretation of neural theory. Fronto-limbic and cognitive-control systems are relevant to threat processing and self-regulation, but a neural correlate does not establish a safe or effective engineering target. Head and hand tracking can document orientation, movement and interaction. It cannot, without extensive validation and a plausible observation model, uniquely identify threat appraisal, motivation or a psychiatric diagnosis. Measurement must be matched to the inferential claim rather than upgraded by terminology such as 'digital biomarker'.

The transition from mechanisms to evidence is therefore not a search for any trial that mentions the same symptom. It asks whether the hypothesized process has been changed in an appropriate population, whether that change matters to functioning and whether simpler methods already achieve the desired effect.

## Evaluate interventions before technology

### Established options set the benchmark

A needs-guided system starts by identifying the strongest feasible alternatives for the same target. CBT has substantial evidence for childhood and adolescent anxiety disorders: the 2020 Cochrane review included 87 studies with 5,964 participants, with clearer evidence relative to no treatment or waiting lists than for some active comparisons and long-term endpoints [11]. This evidence supports appropriate psychological care; it does not imply that any individual CBT component or delivery format is equally effective across diagnoses and settings.

Universal school-based social-emotional learning (SEL) also provides a relevant benchmark. A meta-analysis of follow-up studies identified benefits across multiple outcomes in diverse school populations [12]. Yet the large MYRIAD cluster trial involving 84 schools and 8,376 pupils did not find evidence that adding universal school-based mindfulness was superior to usual social-emotional provision at one year [13]. The lesson is methodological, not that school programmes are unhelpful: a plausible intervention may fail to add value when the comparator already provides active support. Likewise, a synthesis of 18 randomized trials of digital sleep interventions found evidence for some digital CBT for insomnia approaches, while its mean study age of approximately 19 years and variable treatment ingredients limit direct conclusions about younger adolescents [14].

Table 1 compares seven functional targets across established approaches, immersive evidence, comparator adequacy, transfer and technology-selection implications (TABLE 1). The table deliberately does not rank all interventions on one numerical efficacy scale. It preserves differences in age, clinical status, outcome measurement and comparator strength. For a target dominated by school safety, environmental action may be necessary even if a VR rehearsal task improves task performance; for a target dominated by treatable social-evaluative avoidance, structured practice might be suitable across multiple modalities.

### What immersive systems add — and what trials show

Immersive virtual reality (VR) can render interactive three-dimensional environments, vary the timing and proximity of social cues, and record behaviour linked to known task events. Its scientific advantage is the potential to control and replicate contextual conditions while eliciting action under a situated task demand. Research on psychological experiments in VR has established useful regularities in how simulated contexts shape responses, although ecological and construct validity remain distinct questions [15]. These features are genuine capabilities, not yet demonstrations of superior clinical outcomes.

A 2026 meta-analysis of nine randomized trials comprising 929 children and adolescents identified an overall modest signal of benefit for immersive psychotherapy interventions, with a larger estimated effect relative to passive than active controls [16]. The trials combined heterogeneous conditions, doses, active ingredients and outcome instruments; a pooled effect cannot serve as a predicted effect size for classroom participation in the general Chinese adolescent population. A recent synthesis of social-anxiety-related VR exposure similarly highlights promise alongside limited trial size and variation in protocols [17]. A school-anxiety feasibility study with ten adolescents demonstrated the potential to elicit relevant responses but cannot estimate relative effectiveness in the absence of a randomized comparison [18].

Direct active comparisons are particularly informative. In a pilot trial involving 51 adolescents meeting diagnostic criteria for social anxiety disorder, CBT combined with either VR exposure or imaginal exposure reduced symptoms. Some social anxiety outcomes favoured imaginal exposure; fear of negative evaluation did not significantly differ between the two active groups [19]. A subsequent published correction to the article must be considered when extracting its findings [20]. This study shows why improvement within VR should not be equated with an immersive advantage. A trial of VR social-skills training in 90 children with attention-deficit/hyperactivity disorder aged 6–12 years reported benefits on some outcomes relative to comparators [21]; that population and age range are not direct evidence for general 12–18-year-old users. Recent VR and tablet cognitive-training studies similarly require careful separation of trained-task gain from generalization to everyday self-management [22].

Non-immersive digital interventions have a developing evidence base. A 2025 adolescent-focused systematic review documented variation in interventions for anxiety and depression; a meta-analysis of digital support for youth social anxiety identified uncertainty surrounding active ingredients, guidance and study quality[31,32]. Broader syntheses of digital mental health trials and youth digital interventions reinforce the need to compare intervention content, controls and target populations before attributing effects to delivery mode[33,34]. Internet-delivered CBT, guided online modules and school support can provide structured practice without head-mounted displays. An adult trial of automated VR treatment for agoraphobic avoidance demonstrates a developed engineering-to-clinical evaluation pathway, but its indication and model cannot automatically generalize to adolescents[23]. Technology selection asks whether a modality provides a mechanism-relevant capability whose added benefit outweighs equipment, training, supervision, privacy and safety costs.

## Conditional modality selection

We distinguish four broad configurations: in-person or ecological support, non-immersive digital tools, immersive VR, and blended care. Selection follows explicit gates (FIG. 3). First, confirm a youth-relevant functional target and assess plausible mechanisms; if the process is poorly specified, improve assessment rather than escalating technology. Second, establish the current non-immersive benchmark. Third, identify a technical feature—such as precisely controlled evaluative feedback or repeatable embodied rehearsal—that the target plausibly needs and that feasible alternatives cannot provide equally well. Fourth, evaluate anticipated incremental benefit jointly with acceptability, cost, equity and harm. Fifth, define an appropriate active comparison and independent real-world outcome before claiming value.

These gates permit an explicit decision not to use VR. Sleep-related dysfunction may be better served by scheduling changes and CBT for insomnia; peer victimization usually requires safeguarding and school-level action; social-evaluative avoidance may justify an immersive component if the incremental value of controlled social rehearsal is demonstrated. Blended interventions are appropriate when controlled practice is only one part of a wider behavioural and social change process. The point of conditional selection is not technological austerity, but proportional engineering: use only the complexity needed to meet a meaningful goal.

## Engineering a deployable support system

### Modality-flexible reference architecture

A support product is more than a simulated scenario. It includes the people who can access it, how a need is assessed, who guides the session, what practice occurs between sessions, which data are recorded, what happens when difficulties increase, and how outcomes are reviewed. Figure 4 maps these components into a modality-flexible architecture (FIG. 4). Its purpose is to make intervention logic traceable across clinical, educational, behavioural and technical teams; it is not a claim that one commercial device has been validated.

The first layer specifies users and context. Intended users may include typically developing adolescents with mild situational difficulties, young people receiving clinician-supervised treatment, and professionals or caregivers authorized to support implementation. These are distinct routes with different eligibility, contraindications, consent procedures and evidence requirements. Developmental suitability is not an age label alone: literacy, prior experience, physical tolerability, communication needs, disability access and individual preferences matter. Risk indicators requiring professional assessment should trigger referral or a change of service pathway, not automatic escalation of practice intensity.

The second layer defines intervention delivery. For each target, the system should represent the smallest adequate configuration of in-person practice, guided digital modules, immersive scenarios or blended support. A session plan should specify its intended active ingredients, duration or dose, content version, facilitator qualification, between-session exercises and conditions for interruption. Intervention descriptions should be reproducible and sufficiently explicit for replication, in line with reporting principles such as TIDieR [24]. The interface should support meaningful choice and accessible alternatives rather than maximizing time spent engaged with the application.

The third layer provides modular content. A classroom-speaking module can implement graduated participation opportunities and, where immersion is justified, vary audience number, gaze, feedback, proximity and task demands. A peer-interaction module can represent conversation initiation, disagreement, repair and collaborative decision-making under structured constraints. A goal-directed task module might investigate planning and strategy adaptation, but far-transfer claims should remain a research question. The same mechanism may be supported through a video, a role-play worksheet, a clinician-guided task or an immersive scenario. The product specification should map the active ingredient across these choices, enabling valid comparisons rather than confounding technology with intervention content.

### Observability, feedback and adaptive control

A reusable engineering specification should describe each observable at the resolution and reliability required by its intended interpretation. For example, a task event might contain a pseudonymous participant identifier, session and scenario version, condition assignment, event type, synchronized timestamp, completion status and missingness flags. Head and hand six-degree-of-freedom trajectories, controller inputs and gaze data may be logged where available and consented, but they are not necessary for all delivery formats. A data dictionary should separate directly recorded observations from derived features and from interpretive hypotheses.

An adolescent's delayed task initiation can be an observed latency; whether it indicates anxiety, uncertainty, fatigue, low motivation or motor difficulty remains an empirical question. Candidate features should be assessed for within-device accuracy, test–retest reliability, sensitivity to context, convergence with appropriate measures, discriminant validity and measurement invariance across age, setting and accessibility groups. COSMIN guidance on measurement properties provides useful principles for content validity and evaluating claims about instrument performance [25]. Internal improvements in a digital score should not serve simultaneously as the training objective, feedback signal and independent endpoint.

Feedback should communicate actionable, contestable observations: for example, 'you chose to begin the task after two opportunities' rather than 'your social anxiety has fallen by 30%'. Adaptation should be rule-governed and bounded by prespecified constraints. An artificial agent can make interaction more flexible but introduces variability, content and safeguarding challenges; identical experimental conditions may require fixed dialogue structures, content moderation and version control. In youth settings, all adaptations should support pause, refusal and exit without punishment, with documented escalation procedures for distress or unsafe content. Development-specific guidance for extended reality identifies cybersickness, physical safety, privacy and psychosocial risks that require direct design consideration [26].

The fourth layer connects the system back to daily life and governance. Parents, teachers and clinicians have different legitimate roles. Permission to manage device use is not equivalent to blanket access to private disclosures. Data collection should be minimal and tied to a defined research or service purpose; safeguards must be appropriate to local law and institutional oversight. For Chinese deployment, personal information protection and minors' online protection requirements are relevant to data governance, but legal compliance does not substitute for research ethics or evidence of efficacy. Access, equipment availability, supervision burden, accessibility and potential stigma belong in product requirements, not only in limitations discussed after a trial.

### A worked traceability example

Consider an adolescent who wishes to speak in class but repeatedly avoids opportunities because of anticipated negative judgment. Needs assessment establishes the adolescent's goal and the relevant classroom conditions. Mechanism evaluation distinguishes evaluative threat from skill deficits and unsafe peer responses. Established exposure and CBT evidence provides the benchmark; a VR audience scenario is considered only if controlled social feedback offers a practical additional capacity. The product specification then links audience cues, task events, user-controlled progression and independent classroom participation. The primary endpoint is defined outside the VR session, such as the proportion of prespecified, meaningful speaking opportunities voluntarily taken up, with attention to opportunity denominators and the adolescent's preferences. If evaluation identifies active bullying instead, the product configuration must shift to safeguarding and school support. This requirement-level traceability is the central contribution of the reference architecture, not a new sensor or an untested therapeutic claim.

## Validation as a sequence of claims

A technically reliable system can be psychologically invalid; a psychologically interpretable task can fail to change a mechanism; a mechanism can change without producing an improvement in daily life; an effective intervention can be too costly or inequitable to implement. These are different failure modes and demand different research designs. Figure 5 proposes six related evidence gates from needs to adoption (FIG. 5), building on established guidance for complex intervention development and evaluation [27]. Progression should be based on prespecified, context-sensitive criteria rather than one universal technical score.

**Gate 1 — Needs and content validity.** Participatory research with adolescents, caregivers, teachers and clinical professionals clarifies whose objective is represented and how everyday difficulty is operationalized. Agreement on a problem label is insufficient without evidence that the proposed content is relevant and acceptable. Disagreement is useful information about heterogeneous goals. The output is a documented target specification, eligibility definition and candidate outcomes, including a statement of when no individualized digital intervention is appropriate.

**Gate 2 — Delivery fidelity, usability and safety.** Engineering studies verify task timing, display stability, tracking accuracy, data loss, controller and device compatibility, content-version fidelity, interruption procedures, accessibility and adverse experiences. Simulation reliability supports only a delivery claim. A realistic virtual audience cannot compensate for inconsistent scenario timing or an interface that increases distress and excludes users with motion sensitivities. Human-factors testing in schools should evaluate supervision time and the feasibility of consent, cleaning and setup workflows as well as participant comfort.

**Gate 3 — Measurement validity.** Validation should begin by specifying the intended construct and the reference criteria. Test–retest studies, alternative-task studies, convergent and discriminant evidence, and invariance analyses can distinguish genuine stability from device or task artifacts. If a motion-derived 'anxiety index' is not distinguishable from general arousal, task familiarity or head-movement style, the appropriate response is to narrow the interpretation or abandon the index—not to average more correlated features into an opaque composite. Measurement evidence does not establish a therapeutic effect.

**Gate 4 — Mechanism engagement.** Controlled contrasts should test whether a prespecified component changes the proposed process. For social-evaluative avoidance, one can manipulate audience feedback or controllability while measuring threat expectations, safety behaviours and task participation at defined times. Strong mediation claims require careful temporal ordering and attention to shared causes of process and outcome. A change in task performance alone cannot confirm the intended psychological mechanism. A failure at this gate may require new theory or different intervention content rather than a more elaborate interface.

**Gate 5 — Comparative effectiveness.** A clinically or developmentally meaningful endpoint must be prespecified before allocation or analysis. For immersive social practice, an informative counterfactual might be content- and dose-matched guided role-play, imaginal exposure or a non-immersive digital condition. Waiting-list comparisons answer a different question. Trials should identify an estimand and use independent assessors where feasible, appropriate handling of school-level clustering, missing data and multiplicity, and complete adverse-event reporting. CONSORT 2025 and SPIRIT 2025 provide current reporting guidance for randomized trials and protocols [28] [29]. Claims of incremental value should consider precision and an agreed minimum meaningful difference, not statistical significance alone.

**Gate 6 — Transfer, implementation and continued value.** Generalization requires outcomes independent of the training task. For classroom participation, measurement might combine youth goals, prospective opportunity records, blinded observation where feasible and validated scales without treating any one measure as definitive. Follow-up should address durability, cross-context transfer, non-use and possible harms. Implementation studies consider intervention reach, fidelity, cost, supervision, acceptability and equity across schools and communities. Recent work on trialing digital mental-health interventions in everyday contexts reinforces the need for ecologically meaningful causal questions and careful timing of intervention effects [30]. A positive laboratory result cannot by itself justify routine adoption.

Stage gates should allow four evidence-responsive decisions: proceed when claims are supported; refine the target, measurement or component when evidence is incomplete; replace a burdensome modality when simpler support has adequate value; and pause when benefit is unsupported or risks are unacceptable. Different claims require different thresholds, which should be justified by the population, context and intended use. This explicit possibility of abandoning a proposed technological solution is a feature of needs-guided engineering, not a failure of it.

## Research priorities for a needs-led field

The central opportunity is to move from a portfolio of immersive applications to a cumulative science of which problems, mechanisms and delivery features justify engineering investment. Four research priorities follow. First, study the heterogeneity concealed by behavioural labels. Repeated measurements within adolescents, multi-informant contextual assessment and mechanistic perturbations can determine when similar participation problems have different causes. China provides an informative setting for studying school, household and resource contrasts, provided findings are not generalized across the population without appropriate sampling.

Second, evaluate intervention components independently from delivery modality. A factorial design can separate the effect of graduated practice, social feedback, facilitator contact and immersion; otherwise a 'VR versus usual care' study confounds many differences. Third, pursue interpretable measurement rather than maximal sensing. Define the behavioural construct before choosing streams, publish feature definitions and missing-data rules, and validate across tasks, hardware, populations and accessibility conditions. Fourth, test implementation value against the strongest feasible alternative, including circumstances in which conventional educational or psychological support produces comparable outcomes with lower costs.

These priorities define a focused research agenda. A critical synthesis of developmental, intervention and engineering evidence can establish decision principles, whereas the comparative utility of the proposed needs-to-validation framework must be tested empirically. Its rules should be evaluated for their ability to improve the selection and configuration of interventions, resource allocation and real-world outcomes. Future studies should measure not only whether an engineered intervention works, but whether needs-guided selection leads to a better choice than technology-first development.

## Conclusions

An adolescent's need for autonomy, belonging, competence or safety is not a specification for a VR headset. Translating that need into meaningful support requires identifying the functional barrier, testing candidate mechanisms, examining what existing approaches already accomplish, selecting only the technological capabilities that add value, and engineering a service whose outcomes can be independently verified. This needs-to-validation sequence supplies a common language for developmental researchers, clinicians, engineers, schools and families. Its purpose is to make intervention design more discriminating: technological sophistication becomes valuable only when it improves what adolescents can do, choose and experience in everyday life.

## Data availability

No new datasets were generated or analysed for this narrative Review. The published studies discussed in this article are identified in the References.

## Acknowledgements

This work was supported by the Hangzhou Institute of Medicine, Chinese Academy of Sciences (grant no. 2025QD44), and the Joint Fund Project of Tongde Hospital of Zhejiang Province and Hangzhou Institute of Medicine, Chinese Academy of Sciences (grant nos. 2025LHJJ06 and 2025LHJJ14).

OpenAI's ChatGPT was used to assist with selected literature searches and grammatical proofreading. The authors take responsibility for the accuracy and integrity of the manuscript.

## Author contributions

Q.Z. conducted the literature search and evidence synthesis, prepared the figures and tables, and drafted the manuscript. Y.W. contributed to the appraisal and interpretation of clinical evidence on adolescent mental health interventions. T.H. contributed to the evaluation of clinical applicability, implementation considerations and translational implications. N.M. supervised the work, contributed to the development of the theoretical framework, and critically revised the manuscript. All authors reviewed and approved the final manuscript.

## Competing interests

The authors declare no competing interests.

## Table 1 | Evidence for competing approaches

| Functional target | Candidate process | Established options | Immersive evidence / boundary | Modality implication |
|---|---|---|---|---|
| Social-evaluative avoidance | Threat expectancy; avoidance | CBT, guided exposure, imaginal or in-person rehearsal | Controlled social feedback; active-comparator evidence mixed | Only if measured added value outweighs cost |
| Peer communication | Social cognition; practice | Peer practice, social-skills and school support | VR youth clinical subgroups; general adolescent transfer unclear | Targeted adjunct; compare real-peer practice |
| Emotion regulation | Appraisal; regulation | CBT, school SEL and digital support | Modest heterogeneous signals; mechanism evidence incomplete | Prefer established support; test modality effects |
| Attention and goal control | Control; planning | Classroom support, structured strategies | Younger-child/task training evidence; distant transfer uncertain | Research extension, not broad enhancement claim |
| Sleep and stress | Sleep timing; fatigue | CBT-I, routines, school scheduling | No established VR increment for circadian mechanism | Prioritize behavioural and environmental care |
| Problematic digital use | Habit; social ecology | Family, school and behavioural support | No direct established VR-specific adolescent benefit | Avoid immersion by default |
| Bullying and exclusion | Peer norms; safety | Safeguarding; school anti-bullying programmes | Simulated coping cannot replace protection | School-level action first |

CBT, cognitive behavioural therapy; CBT-I, cognitive behavioural therapy for insomnia; SEL, social-emotional learning; VR, virtual reality. Evidence summaries are qualitative and population-specific and are not formal GRADE ratings.

## Box 1 | Chinese contexts as a design stress test

China provides a stringent test of needs-guided engineering because shared developmental needs are expressed through differing school, family, peer and digital circumstances. The same visible withdrawal may reflect anticipated evaluation, limited social opportunity, sleep and stress, reduced agency or peer victimization. Each explanation changes the appropriate target and delivery choice. In one scenario, a young person with evaluation-related classroom avoidance could receive graduated practice, comparing a VR audience with guided non-immersive exposure and evaluating independent classroom participation. In another, avoidance under peer victimization calls first for safeguarding and school-level action rather than more individual rehearsal. Across ages and regions, eligibility, accessible alternatives, privacy, caregiver/professional permissions and costs affect which systems are justified. Context is therefore an effect modifier and a design variable, not a diagnosis or a national psychological profile.

# Figure 1

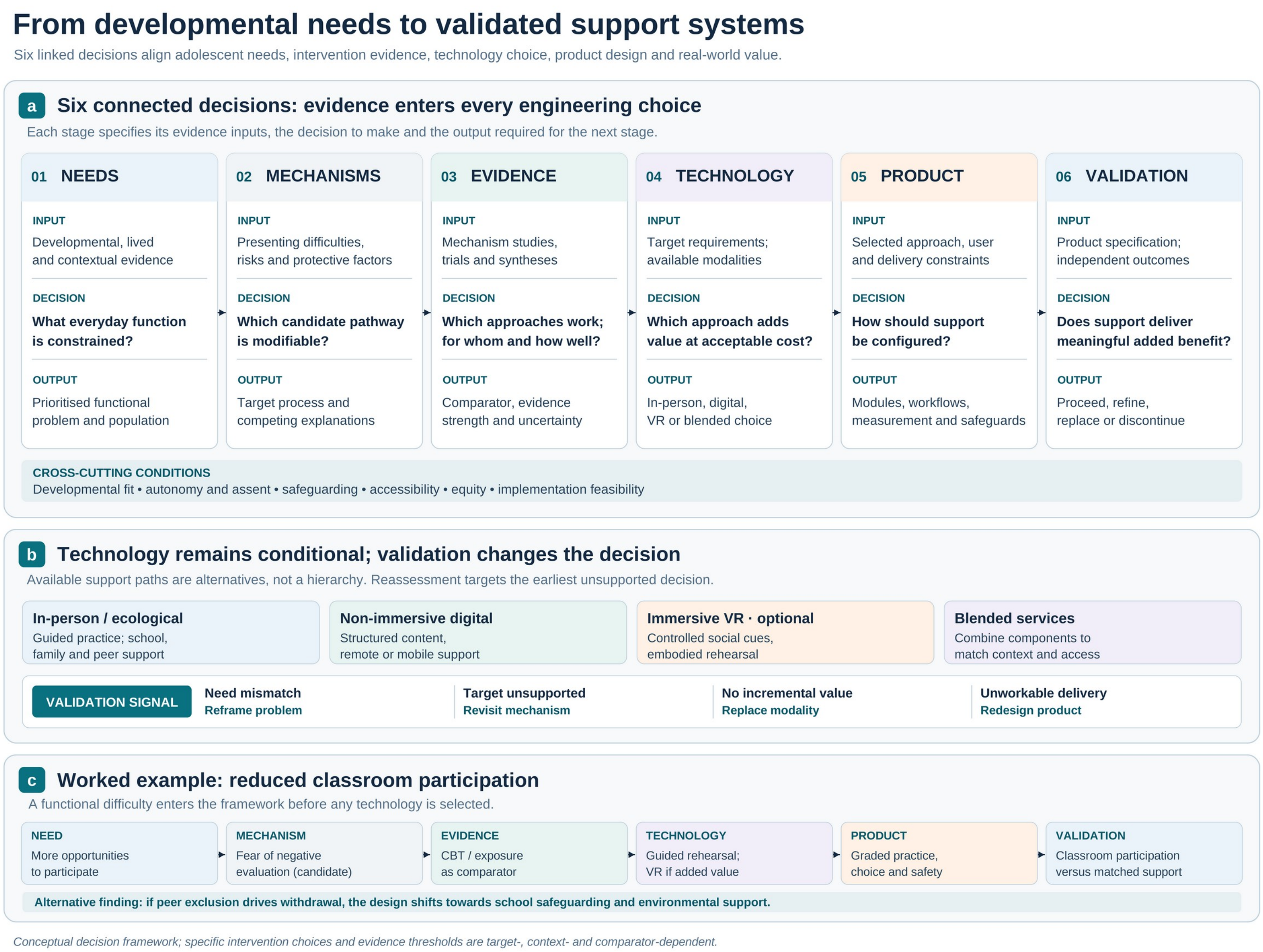

Figure 1 | From developmental needs to validated support systems. (a) Six-stage decision framework connecting adolescent needs, psychological mechanisms, intervention evidence, technology selection, product engineering and validation. Each step makes an inferential claim and yields requirements for the next. (b) Selection among non-technological, non-immersive, immersive and blended delivery, with feedback to revise the earliest unsupported decision. (c) Classroom participation as an illustrative case of traceability from goal to independent outcome. The framework specifies decision principles rather than empirically validated causal pathways.

## Figure 2

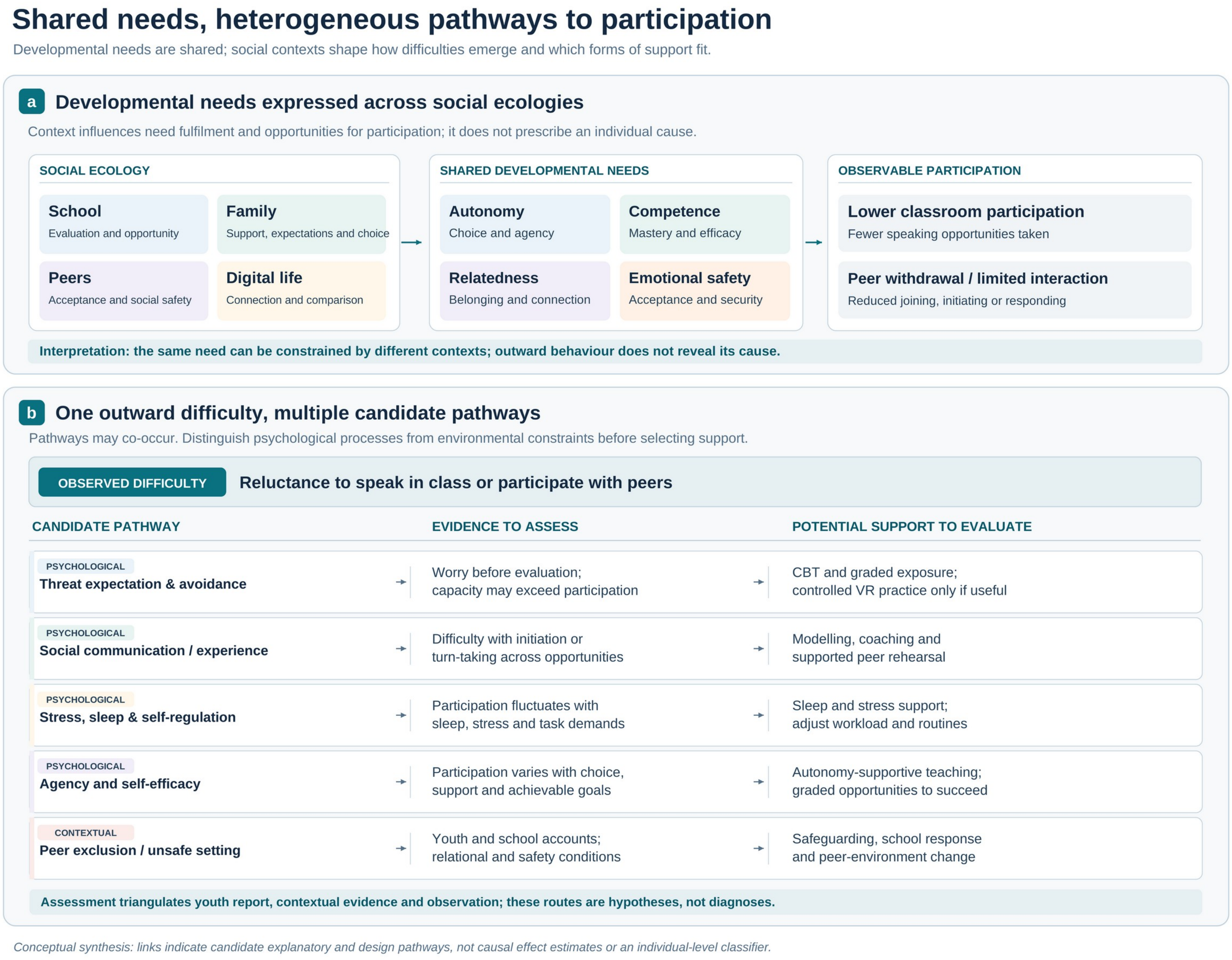

Figure 2 | Shared developmental needs and heterogeneous pathways to participation difficulties. (a) School, family, peer and digital environments shape the opportunities through which autonomy, competence, relatedness and safety are expressed. (b) Similar outward withdrawal may reflect social-evaluative threat, communication experience, stress and sleep, agency or environmental exclusion. Proposed pathways guide assessment and different forms of support, rather than constituting diagnostic categories or quantified effects.

# Figure 3

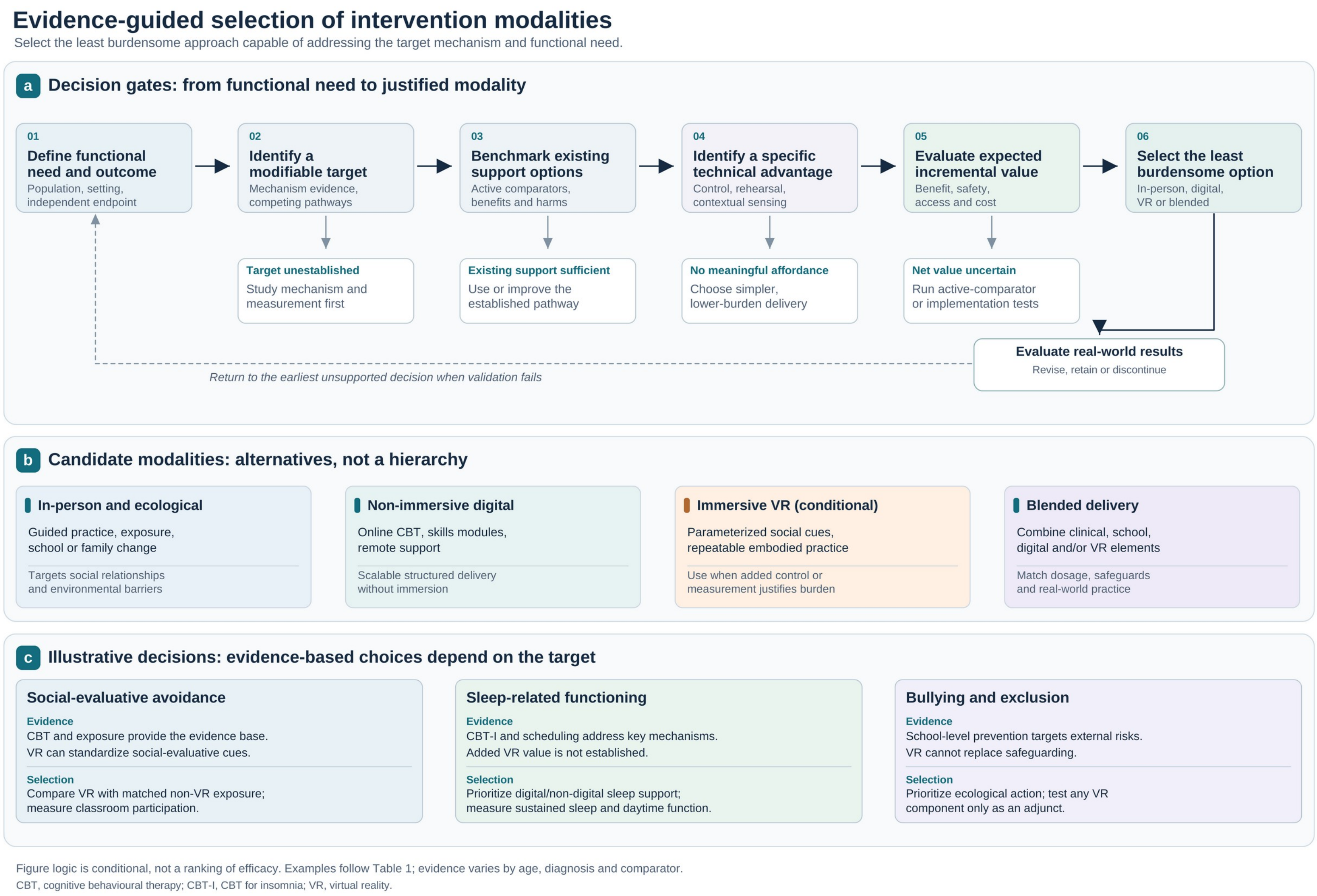

Figure 3 | Evidence-guided selection of intervention modalities. (a) A sequence of conditional gates determines whether the functional target is established, whether existing options suffice, and whether technology offers a justified incremental capability. (b) Feasible modalities include ecological or in-person support, non-immersive digital delivery, conditional immersive VR and blended programmes. (c) Social-evaluative avoidance, sleep problems and peer victimization illustrate contrasting decisions. No efficacy ranking is implied.

# Figure 4

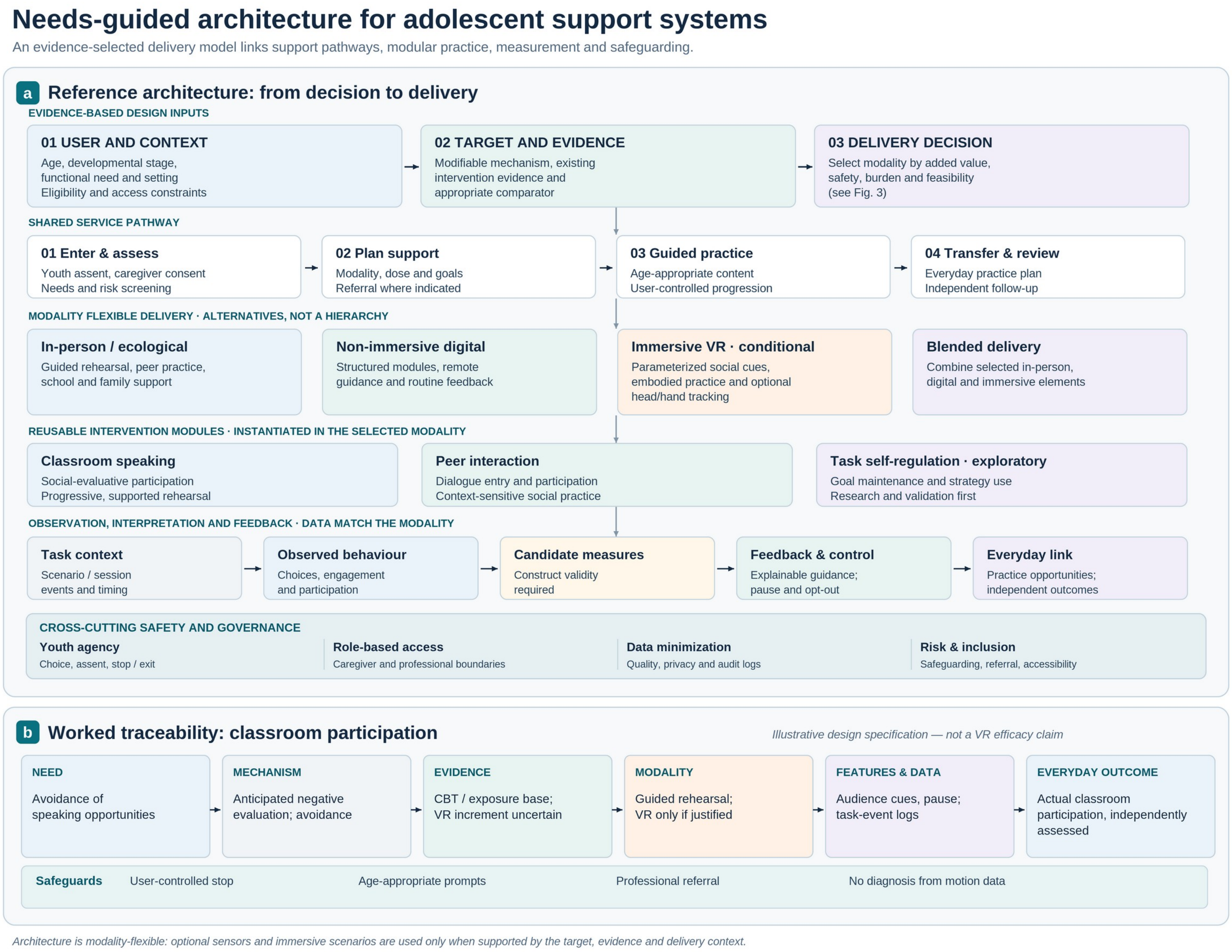

Figure 4 | A needs-guided architecture for adolescent support systems. (a) The architecture links target and evidence definitions to eligibility, service planning, modality-specific modules, event capture, feedback, independent follow-up and safeguards. Data collection is specific to the selected modality. (b) A classroom-participation example traces the need, mechanism, comparator, conditional modality, features and external outcome. Psychological interpretation of observed behaviours requires separate construct validation.

## Figure 5

# Evidence gates from design to real-world adoption

A stage-gated framework tests what a support system delivers, measures, changes and adds beyond existing care.

### a Six evidence gates: advancing from product specification to justified adoption

Each gate supports a distinct claim; success at one gate does not establish success at the next.

**G1 — Needs & content validity**
Does the system target a meaningful difficulty?
**Evidence examples**
Youth co-design; content validity

→

**G2 — Technical fidelity & safety**
Can delivery be reliable, safe and accessible?
**Evidence examples**
Fidelity, usability; adverse events

→

**G3 — Measurement validity**
Do observed features reflect the construct?
**Evidence examples**
Reliability; construct validation

→

**G4 — Mechanism engagement**
Does the intervention change its target?
**Evidence examples**
Component experiments; process measures

→

**G5 — Comparative effectiveness**
Is added benefit meaningful versus a real alternative?
**Evidence examples**
Active-comparator trial; independent outcome

→

**G6 — Transfer & implementation**
Does benefit persist and justify deployment?
**Evidence examples**
Real-world follow-up; cost and equity

**At every gate** Predefine success criteria and harms; document uncertainty; compare with a feasible alternative.

PROCEED evidence adequate · REFINE target / method / product · REPLACE simpler modality · PAUSE insufficient or unsafe

### b Claims, study designs and decisions at each gate

Evidence must support the specific inferential claim; proxy signals and within-task gains are not substitutes for independent outcomes.

| GATE | CLAIM TO ESTABLISH | RESEARCH EVIDENCE / READOUT | IF CRITERIA ARE NOT MET |
|---|---|---|---|
| 1 G1 Need | A relevant, actionable everyday difficulty is specified | Participatory assessment; content validity; school/family context | Reframe the target population, need or primary outcome |
| 2 G2 Delivery | Intervention is reproducible, usable and acceptably safe | Scenario and device fidelity; usability; adverse-event logs | Repair or simplify the modality; do not progress on novelty alone |
| 3 G3 Measure | Behavioural indicators support their stated interpretation | Test–retest; convergent and discriminant validity; invariance | Relabel or drop invalid proxies; avoid psychological-score claims |
| 4 G4 Mechanism | A prespecified modifiable process responds to the component | Controlled component contrasts; temporal process measurements | Revise the target or active ingredient; do not infer mediation from correlation |
| 5 G5 Increment | Functional benefits exceed or justify use vs a feasible comparator | Active-controlled study; prespecified estimand; independent outcomes | Prefer the simpler option unless other meaningful value is shown |
| 6 G6 Adoption | Benefit is durable, equitable and feasible in routine settings | Everyday follow-up; implementation fidelity; access and economic data | Adapt delivery or defer scale-up; test again in the intended setting |

**CROSS-CUTTING REQUIREMENTS** Youth assent and consent · Safeguarding · Data governance · Contextual equity · Transparent reporting

Stages may iterate or run partly in parallel. Adoption requires convergent evidence; threshold values should be prespecified for each target and setting.

Figure 5 | Evidence gates from design to real-world adoption. (a) Needs validity, technical fidelity, measurement validity, mechanism engagement, comparative effectiveness and implementation are separate claims. (b) Appropriate studies and decisions are specified for each gate, with options to proceed, refine, replace or pause. Safeguarding, transparent reporting, access and data governance apply throughout.